\documentclass[12pt,preprint]{aastex}
\def\ltsima{$\; \buildrel < \over \sim \;$}
\def\gtsima{$\; \buildrel > \over \sim \;$}
\def\lsim{\lower.5ex\hbox{\ltsima}}
\def\gsim{\lower.5ex\hbox{\gtsima}}
\def\lapp{\ifmmode\stackrel{<}{_{\sim}}\else$\stackrel{<}{_{\sim}}$\fi}
\def\gapp{\ifmmode\stackrel{>}{_{\sim}}\else$\stackrel{<}{_{\sim}}$\fi}

\newdimen\minuswidth    
\setbox0=\hbox{$-$}
\minuswidth=\wd0
\catcode`@=\active
\def@{\kern\minuswidth}
\newdimen\digitwidth    
\setbox0=\hbox{\rm0}
\digitwidth=\wd0
\catcode`!=\active
\def!{\kern\digitwidth}
 
\shorttitle{Binary stars in the AGB of 47 Tucanae?}  
\shortauthors{Beccari et al.}
 
\begin{document} 
 
\title{A population of binaries in the Asymptotic Giant Branch of 47
Tucanae?\footnote{Based on observations with the NASA/ESA HST, obtained at
the Space Telescope Science Institute, which is operated by AURA, Inc., under
NASA contract NAS5-26555. Also based on WFI observations collected at the
European Southern Observatory, La Silla, Chile, within the observing programs
62.L-0354 and 64.L-0439.}}

\author{Giacomo Beccari\altaffilmark{1,2,3},
Francesco R. Ferraro\altaffilmark{4}, Barbara Lanzoni\altaffilmark{1},
Michele Bellazzini\altaffilmark{1}}
\affil{\altaffilmark{1} INAF--Osservatorio Astronomico 
di Bologna, via Ranzani 1, I--40127 Bologna, Italy, 
giacomo.beccari@bo.astro.it}
\affil{\altaffilmark{2} Dipartimento di Scienze della Comunicazione, 
Universit\`a degli Studi di Teramo, Italy}
\affil{\altaffilmark{3} INAF--Osservatorio Astronomico 
di Collurania, Via Mentore Maggini, I--64100 Teramo, Italy}
\affil{\altaffilmark{4} Dipartimento di Astronomia, Universit\`a 
degli Studi di Bologna, via Ranzani 1, I--40127 Bologna, Italy, 
francesco.ferraro3@unibo.it}
\medskip

\keywords{Globular clusters: individual (47Tuc); stars: evolution
-- binaries: close - blue stragglers}

\begin{abstract}
We have used a set of archived Hubble Space Telescope/ACS images to probe the
evolved populations of the globular cluster 47 Tucanae.  We find an excess of
Asymptotic Giant Branch (AGB) stars in the cluster core. We interpret this
feature as the signature of an extra-population likely made by the progeny of
massive stars originated by the evolution of binary systems. Indeed the
comparison with theoretical tracks suggests that the AGB population of 47 Tuc
can be significantly contaminated by more massive stars currently
experiencing the first ascending Red Giant Branch.
\end{abstract}

\section{INTRODUCTION} 
\label{intro}
In recent years it became clear that stellar evolution and stellar dynamics
cannot be studied independently: in fact, physical interactions between
single stars and binaries, as well as the formation and evolution of binary
systems can significantly alter the properties of the overall stellar
populations. This is particularly true for dense stellar systems such as the
Globular Clusters (GCs), where dynamical encounters between stars (especially
those involving binaries) are most probable.  In particular, such a dynamical
activity can generate {\it exotic} stellar populations, like Blue Straggler
Stars (BSSs), Millisecond Pulsars (MSPs), low-mass X-ray binaries (LMXBs),
Cataclysmic Variables (CVs), etc. \citep[see][]{ba95}, that cannot be
explained by the standard single-mass stellar evolution theory, and are
thought to be originated by the evolution of binary systems.  For instance,
BSSs consist (or will consist, once the mass transfer or the merge event is
completed) in a progeny of mass-enhanced single stars (possibly with a He
white dwarf companion), with masses typically ranging between 1 and 2 times
the cluster Turn Off (TO) star mass.  Then, these objects will evolve along
the characteristic paths of a rejuvenated mass-enhanced population, and are
therefore expected to "contaminate" the post-main sequence (MS) evolutive
sequences of normal cluster stars.  However, when considering all possible
by-products of the binary evolution, a variety of different exotic objects
could be originated (depending on the mass, the evolutionary state of the
donor star, the mass ratio and the physical parameters of the binary system),
and their identifications in the Color Magnitude Diagram (CMD) may be not
obvious.  The detection of surface chemical anomalies also seems to be a
quite promising tool for identifying by-products of binary evolution, but
high-resolution spectroscopic studies of this kind have just started to
become feasible \citep[see][and references therein]{fe06}.


Indeed, 47 Tucanae has been found to harbor a large population of exotic
objects in its core \citep[see][]{g01,fe01,k02}, confirming that stars have
experienced (and are experiencing) dynamical interactions involving single
stars or primordial binaries, leading to mass transfer phenomena and mergers
\citep[see][hereafter F04; Mapelli et al. 2004; Ferraro et
al. 2006]{fe04}. In particular, the large population of BSSs found in the
cluster \citep[][F04]{par91,gu92, fe01} represents the major reservoir of
binary by-products currently experiencing the MS stage.  Their bimodal radial
distribution, discovered by F04, suggests that most of the BSSs strongly
segregated in the cluster center have a collisional origin, while those lying
in its periphery are mainly originated by the evolution of primordial
binaries \citep[see][2006]{ma04}.  Also the detection of chemical anomalies
via high-resolution spectroscopy has recently provided a number of
interesting results in the case of 47 Tuc: {\it (i)} \citet{fe06} have
detected a population of carbon-oxygen depleted BSSs and interpreted this
feature as the signature of the mass-transfer activity during the BSS
formation processes; {\it (ii)} \citet{w06} have detected peculiar {\it s}
and {\it r}-process elements enhancement on the surface of 7 Asymptotic Giant
Branch (AGB) stars in the external region ($r>10'$) of the cluster. They
proposed that at least part of these AGB stars might be formed through mass
transfer along the evolutionary path of a $1.4 + 0.5 M_{\odot}$ binary
system.


%

In this Letter, we further investigate the issue of identifying possible
by-products of binary evolution in 47 Tuc, by reporting on the detection of
anomalies (both in terms of star counts and radial distribution) along the
canonical evolutionary sequences in the cluster CMD. In particular, we have
found a significant enhancement of AGB stars in the core and we discuss the
possibility that this is due to a contamination by a progeny of massive stars
originated by the evolution of binary systems, and now sunk to the cluster center
by mass segregation processes.

\section{Observations and data reduction}
\label{obs}  
In this paper we use high-resolution photometric data obtained with the
Advanced Camera for Survey (ACS) on board the Hubble Space Telescope
(GO-9453). Observations were performed with the {\it ACS-Wide Field Channel},
which employs a mosaic of two $4096\times2048$ pixels CCD, providing a
plate-scale of $0.05\arcsec/pixel$ and a total field of view (FoV) of
$3.4'\times3.4'$.  In these images the core of the cluster is fully contained
in $chip \# 1$ of the ACS CCD mosaic. Since we were interested in deriving an
accurate photometry of the cluster giant population, only images with very
short exposure times ($t_{exp}=0.5$ s) obtained through the filters F606W and
F814W were retrieved from the ESO/ST-ECF Science Archive.  All the images
were properly corrected for geometric distortions and effective flux (over
the pixel area) following the prescriptions of \citet{si05}.  The photometric
analysis was performed independently in the two drizzled images by using the
aperture photometry code SExtractor \citep[{\it Source-Extractor};] []{be96},
and adopting a fixed aperture radius of 2 pixels ($0.1\arcsec$). The two
magnitude lists were cross-correlated in order to obtain a combined catalog.
A sample of bright isolated stars has been used to transform the instrumental
magnitudes to a fixed aperture of $0.5\arcsec$, and the extrapolation to
infinite has been performed by using the values listed in Table 5 of
\citet{si05}.
The magnitudes were finally transformed into the VEGAMAG photometric system
by adopting the synthetic Zero Points from Table 10 of \citet{si05}.


An additional data-set of public short-time exposure ($t_{exp}=20$ s) images
in $V$ and $I$ bands, obtained with the {\it Wide Field Imager} (WFI) mounted
at the 2.2m telescope at ESO, were also retrieved from the ESO/ST-ECF
Archive. The raw WFI images were corrected for bias and flat field, and the
over-scan region was trimmed using standard IRAF\footnote{IRAF is distributed
by the National Optical Astronomy Observatory, which is operated by the
Association of Universities for Research in Astronomy, Inc., under
cooperative agreement with the National Science Foundation.} tools.  The PSF
fitting procedure was performed independently on each $V$ and $I$ image,
using DoPhot \citep{dophot}.

Each WFI catalog was referred to the absolute astrometric system by adopting
the procedure described in F04. The astrometric Guide Star Catalog GSCII was
used to identify astrometric standards in the WFI FoV.  More than thousand
stars were used to find an astrometric solution for each of the eight WFI
chips, with an accuracy of the order of $\sim0.2\arcsec$ in both right
ascension and declination.  Then, we used $\sim1000$ stars in common between
WFI and ACS catalogs as {\it secondary astrometric standards} to transform
the ACS catalog into the WFI astrometric system. Since the two data sets have
the $I$ passband in common (F814W filter corresponds to the $I$ band), we
used the stars in common to photometrically homogenize the two samples.

In order to avoid spurious effects due to incompleteness of the ground based
observations in the most crowded region of the cluster, we restricted the WFI
dataset to the outer region ($r>224\arcsec$; hereafter the {\sl WFI
sample}). In the inner region only stars observed with ACS have been
considered. However, since the ACS FoV was not centered on the cluster
center, the most external region ($r\sim 170-224\arcsec$) was only marginally
sampled by the high-resolution observations; for this reason we decided to
exclude it from the analysis. Hence in the following we consider in the {\sl
ACS sample} only stars with $r<170\arcsec$ from the cluster center.


\section{RESULTS}
\label{sele}
\begin{figure}
\plotone{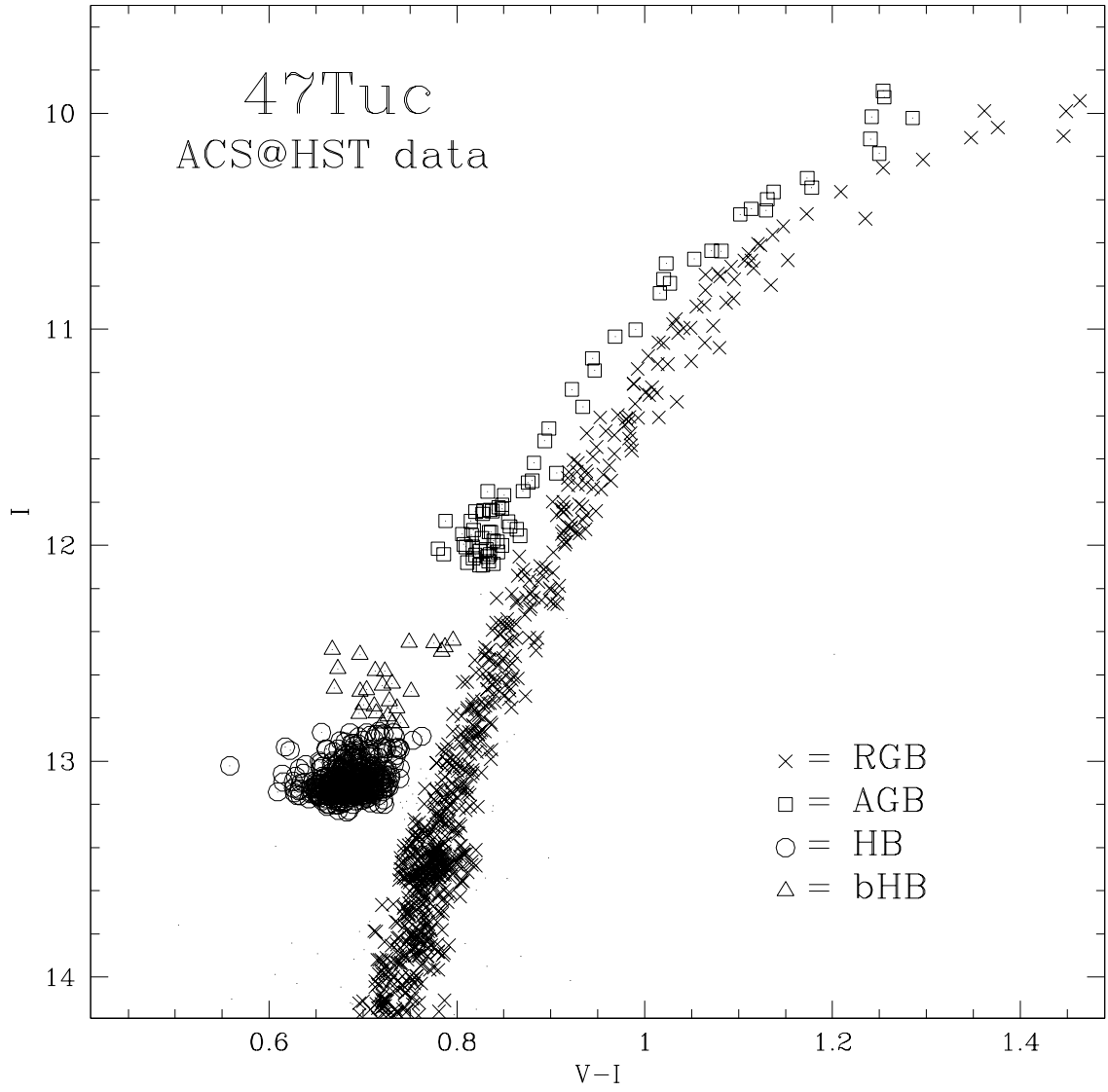} 
\caption{\label{cmd} \footnotesize{CMD for the ACS sample. Stars marked with
different symbols belong to different populations }}
\end{figure}

In Figure~\ref{cmd} we show the brightest portion of the CMD obtained from
the analysis of the ACS sample.  Thanks to the high quality of the data, all
the sequences are well populated and clearly separable one from the others.
In particular the AGB-clump is clearly visible at $I\sim 12$. This
feature, which flags the beginning of the AGB evolutionary phase, has been
observed in other clusters (Ferraro et al. 1999) and it is well predicted by
theoretical models (e.g., Castellani, Chieffi \& Pulone 1991).  In fact the
beginning of the AGB phase is characterized by a slowing down of the stellar
evolutionary rate, just after the very fast evolution at the end of the
Horizontal Branch (HB) phase \citep[see Figure 1 by][]{cast91}. This is
nicely confirmed by the star distribution in the CMD shown in
Figure~\ref{cmd}, where only a few stars can be found between the HB-clump at
$I \sim 13.1$ and the AGB-clump. We therefore selected samples of
stars in different evolutionary stages as marked in Figure~\ref{cmd} by
different symbols: Red Giant Branch (RGB; {\it crosses}), HB\footnote{We
considered ``fiducial HB'' only stars belonging to the HB-clump, while those
lying in the HB extension toward brighter magnitudes have been considered as
a different population and labeled as {\it bright}-HB (hereafter bHB) stars
(they are marked with {\it triangles} in Figure 1), since they can be 
sensibly contaminated by a progeny of binaries
currently experiencing the Helium-burning phase (see the discussion in
Section 4).}  {\it (open circles)}, and AGB {\it (open squares)},
respectively.

In {\it panel (a)} of Figure 2 we show the cumulative radial distribution of
the selected populations in the innermost region of the cluster
($r<100\arcsec$).  Surprisingly AGB stars turn out to be significantly more
centrally segregated than HB and RGB stars. In order to check the statistical
significance of this result we applied a Kolmorogov-Smirnov 
test, that yields a $\sim99\%$ probability that the radial distribution of
AGB stars is genuinely different from that of both HB and RGB stars. An
analogous test has been performed for the WFI sample, and no significant
difference in the radial distribution of the three populations has been
detected in the outer regions of the cluster.


In order to further investigate this unexpected result, we followed the
approach proposed by F04 for the study of BSSs in this cluster: we divided
the surveyed area (ACS and the WFI FoVs) into a set of 9 concentric annuli.
Thus, HB and AGB stars has been counted in each annulus (see Table 1) and the
population ratio $N_{AGB}/N_{HB}$ have been computed.
Accordingly with the definition by \citet{fe93}, we also computed the {\it
double normalized} specific frequency $R_{\rm AGB}$\footnote{Note that this
ratio is expected to be equal to unity for any post-MS evolutionary stage,
since the observed number of stars is expected to scale linearly with the
sampled luminosity following equation 1 of \citet{rf88}.}:
\begin{displaymath}
R_{\rm AGB} = {{(N_{\rm AGB}/N_{\rm AGB}^{\rm tot})} \over 
{(L^{\rm sampled}/L_{\rm tot}^{\rm sampled})}} 
\end{displaymath}
where the fraction of AGB stars observed in each annulus ${(N_{\rm
AGB}/N_{\rm AGB}^{\rm tot})}$ is normalized to the fraction of luminosity
sampled by each annulus ${(L^{\rm sample}/L_{\rm tot}^{\rm sample})}$, as
computed from the surface brightness profile.
Figure~\ref{rad_acs} shows the radial distribution of the calculated ratios.
As can be seen, AGB stars turn out to be significantly overabundant
in the innermost annulus ($r<20\arcsec$, roughly corresponding to 47 Tuc core
radius): in fact, while in the outer regions ($r>20\arcsec$) the
$N_{AGB}/N_{HB}$\footnote{Note that the same result has been obtained by
using the RGB as reference population.} ratio is roughly constant with a mean
value ($0.13\pm0.03$) which is in full agreement with what expected on the
basis of the evolutive time scales~\citep[see][for a review]{rf88}, the same
ratio in the cluster core is $\sim 0.30\pm0.06$, suggesting that AGB stars
are twice more abundant there than in the external regions.  Also the
$R_{AGB}$ ratio, which turns out to be $\sim 1.5$ in the central bin,
suggests the presence of $\sim 10$ contaminating stars (mimicking low-mass
AGB).


\section{DISCUSSION}

We have found evidences of a significant overabundance of AGB stars in the
core of 47 Tuc. This finding is in agreement with the preliminary detection
of an {\it "enhancement in the frequency of stars about 1 mag brighter than
the HB"} already noted by \citet{ba94}.  As shown in Figure 3 this
enhancement is present only in the cluster core.  Such a strong radial
segregation is not expected to result from dynamical processes (mass
segregation) acting on genuine AGB stars\footnote{AGB stars are expected to
be even less massive than the current TO stars because they have experienced
significant mass-loss during previous evolutionary stages.}. Hence the
combination of these observational facts (overabundance and central
segregation) suggests the presence along the AGB of 47 Tuc of an {\it
extra-population} of massive (not-genuine AGB) stars, possibly related to the
evolution of binary systems, i.e., generated either by mass transfer (or
coalescence) between binary companions, or by mergers of single/binary stars
due to collisions. Indeed previous multiband observations \citep{cam00, g01,
fe01, k02} have demonstrated that the core of 47 Tuc hosts a large population
of centrally segregated exotic objects (BSSs, LMXBs, CVs, MSPs, etc,), that
can be originated by a variety of interacting binaries. In particular, the
large population of BSSs, having masses between 1 and 1.6 $M_{\odot}$, are
the natural progenitors of any anomalous high-mass evolved object that can be
currently found in the cluster. In this respect, \citet{ba94} first suggested
that (at least part of) the excess AGB stars might be evolved BSSs currently
experiencing the HB evolutionary stage.  However an appropriate comparison
with theoretical tracks (see Figure 4) clearly demonstrates that, in a metal
rich cluster as 47 Tuc, the Zero Age Horizontal Branch (ZAHB) magnitude level
of stars in the mass range $1.1-1.5 M_{\odot}$ is significantly fainter than
the observed AGB-clump.  Instead, the CMD region populated by the lower mass
AGB is fully consistent with the RGB sequence of such massive stars: hence,
the observed AGB population can be significantly contaminated by high-mass
binary by-products currently ascending for the first time the RGB. Also the
evolutionary time-scales suggest that this could be the case: in fact, a $1.3
M_{\odot}$ star spends $\sim2.4\times10^7$ yr evolving along the RGB in the
magnitude portion covered by the low-mass AGB extension ($12.2<I<10.2$).
This time-scale is more than twice the AGB evolutionary time ($\sim 10^7$ yr)
of a genuine $0.8 M_{\odot}$ star.

From the inspection of Figure 4 we also note that the ZAHB level of a
$1.1-1.5 M_{\odot}$ star is fully consistent with the brightest extension of
the observed HB-clump (i.e., stars labeled as bHB and plotted as empty
triangles in Figure 1). Hence (at least part) of the bHB stars observed
between the HB and the AGB clumps could be the progeny of the binary systems
currently experiencing the He-burning phase. Interesting enough the radial
distribution of these stars turns out to be significantly different (at more
than $3\sigma$ level) from that of genuine HB (or RGB) stars, and similar to
that of AGB stars (panel $b$ of Figure 2).  To further investigate the
possibility that the canonical sequences are contaminated by binary
populations, we studied the radial distribution of stars at the base of the
RGB. In particular, we found that $\sim 30$ faint ($13.7< I <14.7$) blue
($V-I \lsim 0.7$) RGB stars (fRGB) show a radial distribution which is
significantly different (at $3\sigma$ level) from that of the other RGB
within the same magnitude limits, and nicely similar to that of AGB and bHB
stars (panel $b$ of Figure 2).  Moreover, the radial distribution of AGB, bHB
and fRGB stars in the innermost region of 47 Tuc nicely agrees with that of
BSSs \citep[][2004]{fe01}.  This seems to indicate a common evolutionary
scenario for these four populations\footnote{Note that ratio of the presumed
contaminating population in the upper RGB (mimicking the AGB) and in the fRGB
is $\sim 3$, in reasonable agreement with the expected evolutionary
time-scale ratio ($\sim 4$).}, suggesting that all these objects have been
segregated in the cluster core by dynamical processes.  As demonstrated by
appropriate dynamical simulations computed to model the BSS peculiar radial
distribution \citep[][2006]{ma04}, dynamical friction processes can generate
the observed central peak of the BSS distribution.  Accordingly to these
findings, the dynamical friction time-scale here specifically computed for 47
Tuc (by using an appropriate model of the cluster) suggests that all the
objects with masses of $1.2-1.5 M_{\odot}$ generated within $\sim 10 r_c$
(corresponding to $\sim 4'$) from the cluster center are expected to sink
into the cluster core in a time comparable to the cluster age ($t=12 Gyr$);
hence they could generate the segregated populations detected in the core.

The evidences reported in this Letter show the presence of an extra
population contaminating the genuine AGB stars in 47 Tuc: these objects are
significantly more centrally segregated than RGB and HB stars suggesting they
are significantly more massive than the normal stars in the cluster.  Hence,
they possibly are the progeny of binary system evolution.  Future
high-resolution spectroscopic observations aimed to search for chemical
anomalies (as those observed in 7 AGB stars in the outer region of the
cluster and in a sub-sample of BSS by Wylie et al. 2006 and Ferraro et
al. 2006, respectively), would identify the binary progeny and possibly
confirm the scenario presented here.


\acknowledgements{} We acknowledge the anonymous referee for useful comments.
This research was supported by contract ASI-INAF I/023/05/0, PRIN/INAF 2006,
and the Ministero dell'Istruzione, dell'Universit\`a e della Ricerca.

\newpage
\begin{deluxetable}{rrr}
\scriptsize \tablewidth{6.5cm}  
\tablecaption{\label{rad} Number of AGB and HB stars in concentric annuli at 
different distances from the cluster center.
}
\startdata \\
\hline
\hline
 & & \\
 $r[arcsec]$      &  $N_{AGB}$	&   $N_{HB}$ \\
 & & \\
\hline        
 & & \\
   0--21 	&  30  &  101    \\
  21--58 	&  23  &  195    \\
  58--110	&  15  &  104    \\
 110--224	&   6  &   45    \\
 224--350	&  19  &  145    \\
 350--550	&  18  &  152    \\
 550--730	&   8  &   61    \\
 730--1000	&   5  &   30    \\
\hline
\enddata
\end{deluxetable}

\begin{figure}
\plotone{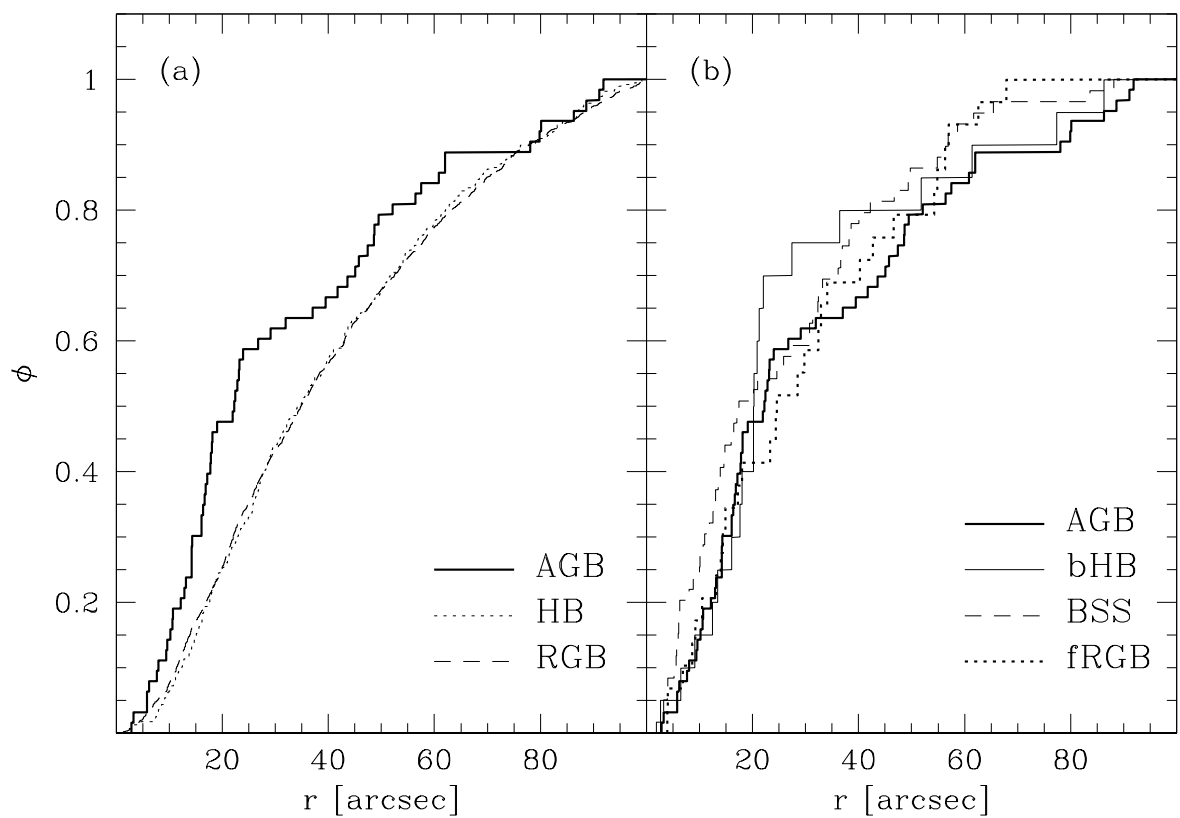} 
\caption{\label{acs_50} \footnotesize{ The cumulative radial distribution of
AGB stars ({\it heavy solid line}) in the innermost region of the cluster
($r< 100 \arcsec$) as a function of the projected distance r from the cluster
center, is compared with those obtained for HB and RGB stars ({\it panel a}),
and bHB, faint-blue RGB (fRGB), and BSS stars ({\it panel b}),
respectively. AGB stars are significantly more centrally segregated than RGB
and HB stars. The radial distribution of AGB, bHB and fRGB stars is nicely in
agreement with that of BSS stars.  }}
\end{figure}

\begin{figure}
\plotone{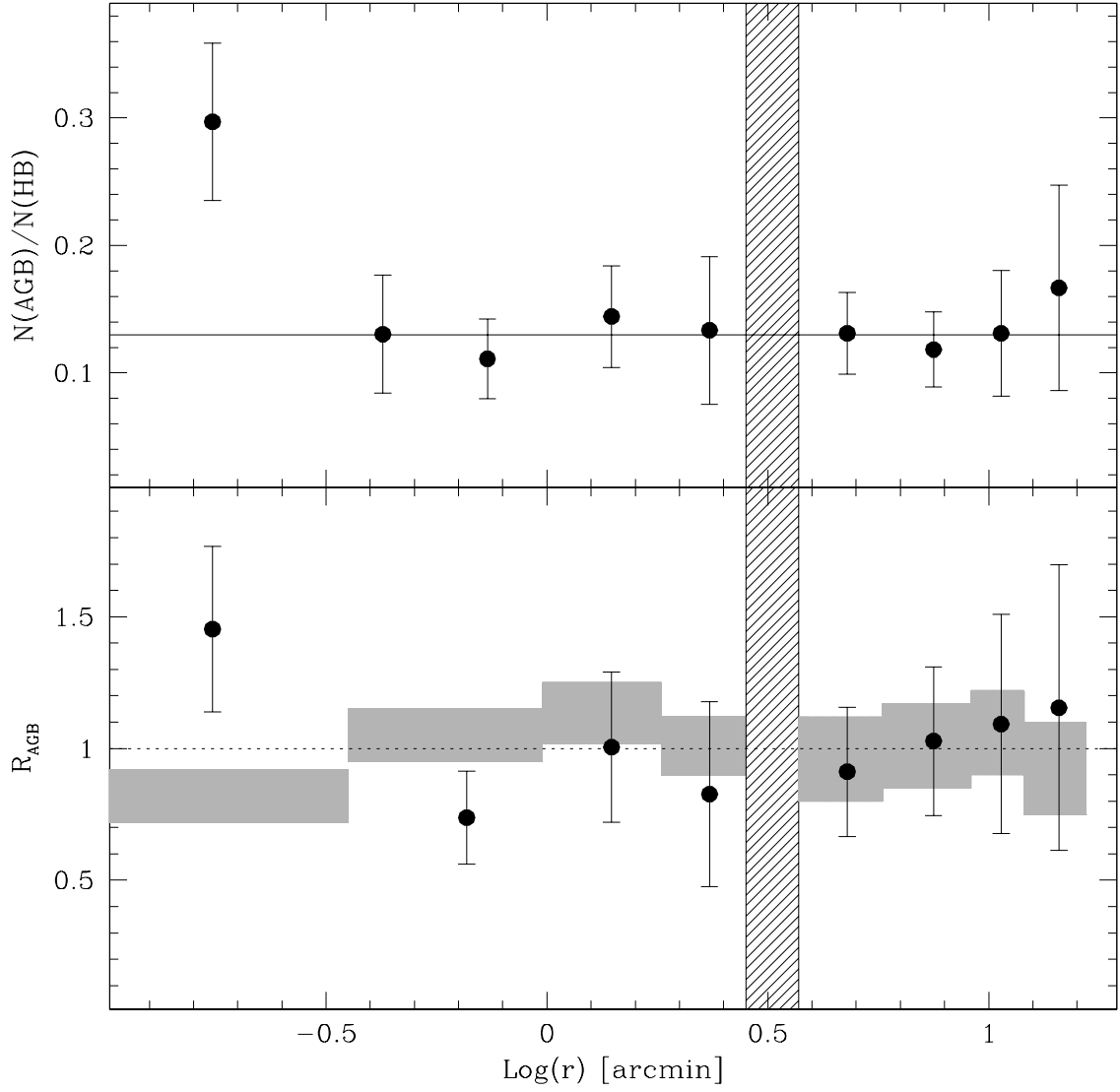} 
\caption{\label{rad_acs} \footnotesize{Relative frequency of AGB-to-HB stars
{\it (upper panel)} and double-normalized specific frequency of the AGB stars
{\it (lower panel)} as a function of the projected distance from the center,
over the entire cluster extention.  Horizontal grey regions in the {\it lower
panel} show the double-normalized specific frequency of HB stars used as
reference population.  The dashed area marks the cluster region excluded form
the analysis in order to avoid spurious effects due to inclompleteness of the
sample. }}
\end{figure}

\begin{figure}
\epsscale{1.1}
\plotone{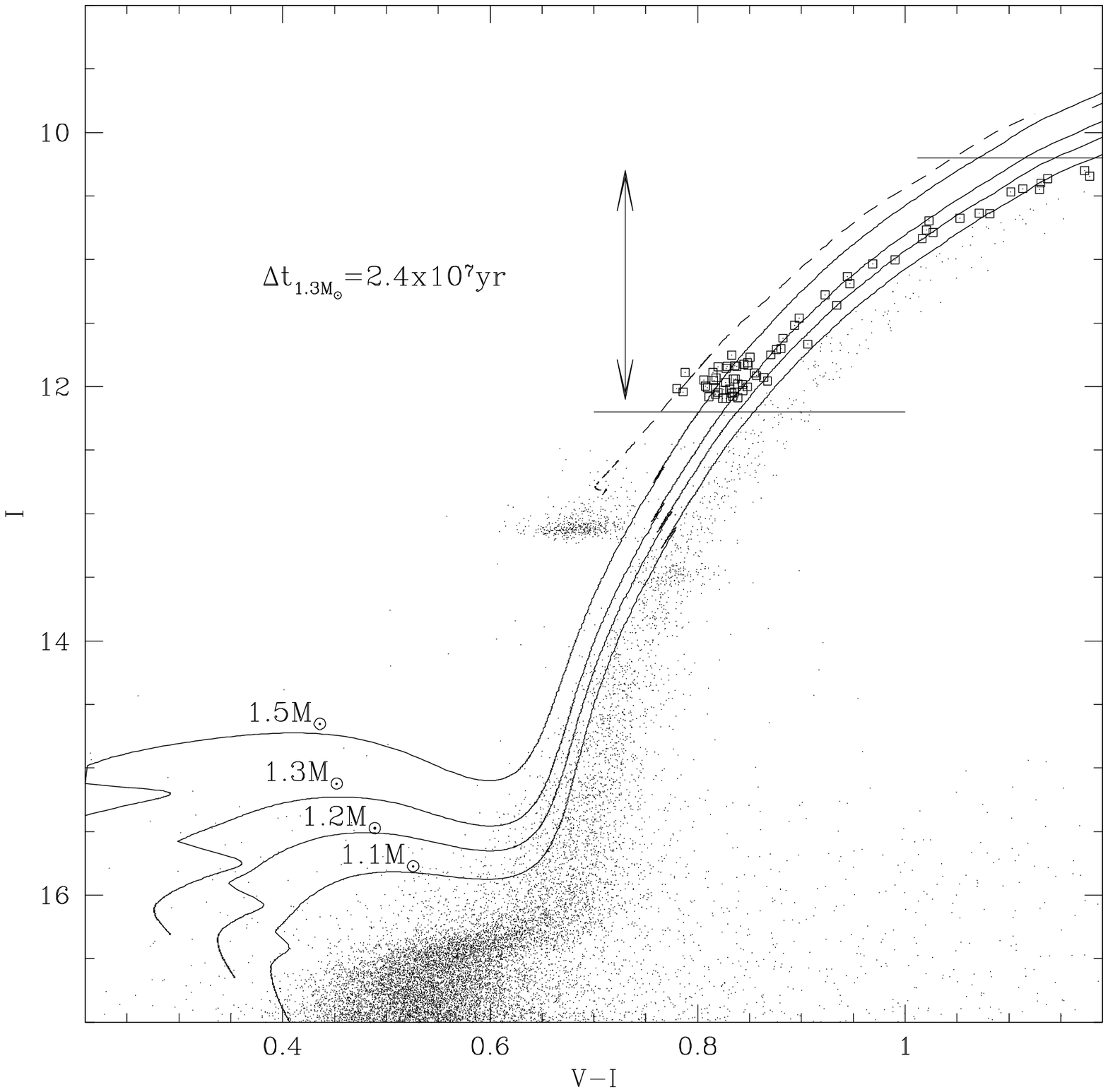} 
\caption{\label{track} \footnotesize{Theoretical tracks from
\citet[][http://www.te.astro.it/BASTI/index.php]{p06} for different masses
(from 1.1 to $1.5 M_{\odot}$; solid lines) are overplotted to the ACS CMD of
47 Tuc.  The theoretical tracks, with $\alpha-enhancement$ ([$\alpha$/Fe]= 0.4)
and [Fe/H]=-0.66 (Z= 0.008), have been transformed in the ACS filters
following \citet{bed05}.  A distance modulus (m-M$)_0=13.33$ and reddening
E(B-V)=0.04, have been adopted \citep{fe99}.  The RGB phase for such massive
stars turns out to occupy the same region of the CMD where "normal" low-mass
AGB stars are observed.  The post-RGB evolutionary track for a $1.5
M_{\odot}$ star is also shown (dashed line).  Note that the ZAHB level for
such a massive star lies in the bright extension of the observed HB where the
bHB sample has been defined.  The evolutionary time needed by a $1.3
M_{\odot}$ star to cover the RGB portion between the two horizontal lines is
also reported. }}
\end{figure}

 
\end{document}